# Designing an Efficient Delay Sensitive Routing Metric for IEEE 802.16 Mesh Networks

*Ishita Bhakta*
School of Mobile Computing and Communication
Jadavpur University,
Kolkata, India
e-mail : ishita.official@gmail.com

*Sandip Chakraborty*
Department of Computer Science and Engineering
Indian Institute of Technology Guwahati,
Assam, India
e-mail : c.sandip@iitg.ernet.in

*Barsha Mitra , Debarshi Kumar Sanyal ,
Samiran Chattopadhyay*
Department of Information Technology
Jadavpur University,
Kolkata , India
e-mail : mitra.barsha@gmail.com,
debarshisanyal@gmail.com, samiranc@it.jusl.ac.in

*Matangini Chattopadhyay*
School of Education Technology
Jadavpur University
Kolkata,India
e-mail : chttpdhy@yahoo.com

*Abstract*—Quality of Service provisioning is one of the major design goals of IEEE 802.16 mesh networks. In order to provide quality delivery of delay sensitive services such as voice, video etc., it is required to route such traffic over a minimum delay path. In this paper we propose a routing metric for delay sensitive services in IEEE 802.16 mesh networks. We design a new cross layer routing metric, namely Expected Scheduler Delay (ESD), based on HoldOff exponent and the current load at each node of the network. This proposed metric takes into account the expected theoretical end-to-end delay of routing paths as well as network congestion to find the best suited path. We propose an efficient distributed scheme to calculate ESD and route the packets using source routing mechanism based on ESD. The simulation results demonstrate that our metric achieves reduced delay compared to a standard scheme used in IEEE 802.16 mesh, that takes hop count to find the path.

*Keywords - IEEE 802.16; mesh; QoS; Minimum Delay; ESD*

## I. INTRODUCTION

The IEEE 802.16 standard, known as "Air Interface for Fixed Broadband Wireless Access Systems" [1] is the next generation *Wireless Metropolitan Area Network (WMAN)* that targets at providing last mile wireless broadband access. The IEEE 802.16 technology can operate in two modes - *Point to Multi point (PMP) and Mesh mode*. While the standard currently defines five classes of services for PMP mode, but there is no QoS provisioning mechanism mentioned in mesh mode. Most of the QoS requirements take delay as an important parameter for providing quality delivery of traffic. The traditional 802.16 mesh network uses standard Dijkstra's or Bellman-Ford's shortest path algorithm which is not suitable for delay sensitive services.

There are many routing metrics available in the context of IEEE 802.11 multi-hop ad-hoc networks, such as ETX, ETT and ML. These metrics are computed using the handshaking property (DATA-ACK) of IEEE 802.11 DCF based MAC. The scheduling mechanism in IEEE 802.16 uses an election based scheme for contention resolution, which is different from 802.11 DCF based contention resolution technique. So the ACK based delay or loss computation is not applicable to 802.16 mesh network. Furthermore, the deterministic election based scheduling in IEEE 802.16 cooperative mesh mode introduces extra scheduler delay at each node, in terms of HoldOff time and contention time. So cross layer properties such as MAC layer delay and routing delay over multiple hops play an important role in routing decision. Most of the current works in IEEE 802.16 consider interference based optimization to find the proper routing decision, but they do not use cross layer inputs to find the path between the source and the destination.

In this paper we propose a cross layer routing metric to provide quality delivery of delay sensitive services. Our scheme is based on an important parameter used in the MAC layer scheduling of IEEE 802.16 Mesh Distributed Coordinated Scheduler, called the *HoldOff Exponent*. In [2], Cao *et al.* proposed a mathematical model for performance evaluation of IEEE 802.16 mesh distributed coordinated scheduler. From the mathematical model proposed by them, we see that a suitably chosen HoldOff exponent can improve the scheduler performance dramatically by minimizing the number of waiting slots between two transmissions. We use this mathematical expectation to find out the minimum delay path between a source and a destination. We also consider the congestion at a network node while designing the metric. Hence this delay-aware cross-layer routing metric is quite suitable for next-generation wireless mesh networks. Simulation results show that our proposed metric works better than standard routing algorithm that takes hop count as a routing metric.

## II. RELATED WORKS

Several works exist in literature for QoS provisioning in IEEE 802.16 PMP mode, but very few of them are for mesh mode. In [2], the authors calculated a fixed point iterative equation for mathematical expectation of the number of slots between two successful control message (MSH-DSCH message) transmissions, and showed that the data transmission and the overall performance of the system depends on control message scheduling. Based on this model, Bayer *et al.* [3] proposed a node differentiation scheme using the effect of HoldOff exponents on the scheduler performance. But this scheme requires excessive analysis of each node's traffic, and the convergence time is high. In [4], Wang *et al.* derive a dynamic HoldOff exponent adjustment algorithm based on the number of two-hop neighbors and the data traffic analysis. In [5], the authors proposed a scheme for tuning HoldOff exponents to minimize the delay at the network. The QoS provisioning mechanisms for IEEE 802.16 mesh network have been studied in [6] and [7].

All the above works are on scheduler performance. Li *et al.* [8] consider routing for minimizing the end-to-end delay for delay sensitive services. Their approach selects the next hop dynamically, using a greedy approach. In [9], Tsai *et al.* introduce a routing and admission control mechanism in IEEE 802.16 mesh distributed network. Their scheme is based on the bandwidth estimation and token bucket admission control.

A review of existing literature on wireless mesh networks such as [10] and [11] reveal that a number of routing metrics are available for wireless mesh networks based on IEEE 802.11 networks. Some of these are - Hop Count (HC), Expected Transmission Count (ETX), Expected Transmission Time (ETT) and Minimum Loss (ML). HC considers the effect of path lengths on the performance of flows choosing the path containing the least number of hops. ETX is one of the first metrics specifically proposed for IEEE 802.11 ad-hoc networks. The ETX of a wireless link represents the expected number of data transmissions required to send a packet over that link to a neighboring node, including retransmissions. The ETX of a route is the sum of the ETX of each link in the route. So ETX selects the route having the highest probability of packet delivery. ETT is an improvement over ETX that takes into account the differences in transmission rates of the links. The ETT of a link *l* is defined as the expected time required to successfully transmitting a packet over a link *l*. The weight of a path is the summation of the ETT's of the links along that path. ML finds the route where the probability of end-to-end packet loss is minimal. It multiplies the delivery ratios of the links in the reverse and forward directions to select the most suitable path.

The above mentioned metrics, namely HC, ETX, ETT and ML were mainly designed for routing in IEEE 802.11 networks and they calculate the required parameters mainly using the ACK frames used for DCF handshaking. The DATA-ACK based handshaking is not used in IEEE 802.16 mesh cooperative scheduling. As described earlier, IEEE 802.16 mesh scheduler introduces extra scheduling delay as computed in [2]. Furthermore, congestion effect is also required to consider in large network scenarios. So, the above mentioned metrics are not directly applicable for IEEE 802.16 mesh network. In this paper, we propose a new cross layer based routing metric, namely *Expected Scheduler Delay (ESD)* that takes scheduler delay and congestion into account. A comparison of ESD with the other metrics is shown in Table 1.

TABLE 1.  COMPARISON OF ROUTING METRICS

| Characteristics | HC | ETX | ETT | ML | ESD |
|---|---|---|---|---|---|
| Quality-Aware | NO | YES | YES | YES | YES |
| Cross-Layer Support | NO | NO | NO | NO | YES |
| Link-load consideration | NO | NO | NO | NO | YES |
| Delay consideration | NO | NO | NO | NO | YES |
| Congestion Avoidance | NO | NO | NO | NO | YES |

### III. IEEE 802.16 MESH SCHEDULING

In IEEE 802.16 mesh distributed coordinated scheduling MSH-DSCH messages are used for transmitting scheduling parameters. MSH-DSCH control messages carry a special parameter called Transmit HoldOff Exponent (*TxtHoldOffExp*) that determines the number of transmission opportunities a node must wait before next contention, after a successful transmission. The number of such transmission opportunities is called HoldOff time given by the following equation,

$$HoldOffTime = 2^{TxtHoldOffExp+Base} \qquad (1)$$

The base value is set to 4 according to the standard to guarantee minimum requirements of fairness.

After hold-off time, each node has to go through a *pseudo-random election algorithm* to find out the next transmission opportunity where the node is eligible to transmit a control message. In [2], Cao *et al.* propose a mathematical model for this pseudo-random election algorithm and calculate the expected number of slots between two successful transmissions as follows,

$$E[\tau] = 2^{x+Base} + E[S] \qquad (2)$$

where E[S] is the expected number of contention slots, x is its TxtHoldOffExp and E[τ] is the number of transmission opportunities between two successful transmissions, which is equal to the summation of HoldOff time and contention time. $E[S_k]$ for node k is given by following equation,

$$E[S_k] = \sum_{j=1,\ j \neq k,\ x_j \geq x_k}^{\aleph_k^{known}} \frac{2^{x_j} + E[S_k]}{2^{x_j+base_j} + E[S_j]} +$$
$$(\sum_{j=1,\ j \neq k,\ x_j < x_k}^{\aleph_k^{known}} 1) + \aleph_k^{unknown} + 1, where$$
$$k = 1, \ldots\ldots\ldots, N$$

(3)

Here $\aleph_k$ denotes the set of 2-hop neighbor nodes of node k. $\aleph_k^{unknown}$ is the set of nodes whose scheduling information are unknown in the neighbor nodes set $\aleph_k$. Let $\aleph_k^{known} = \aleph_k \setminus \aleph_k^{unknown}$. The above equation determines the expected number of slots between two successful transmissions and can be solved using a fixed point iteration.

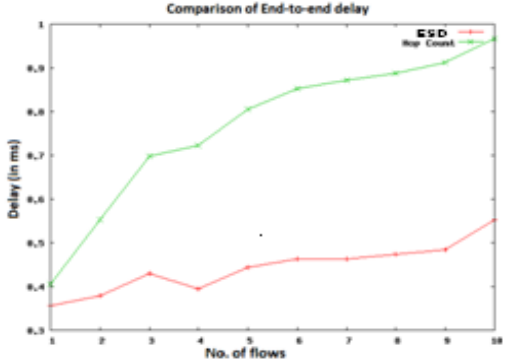

Figure 1. Delay Comparison with HoldOffExp = 0

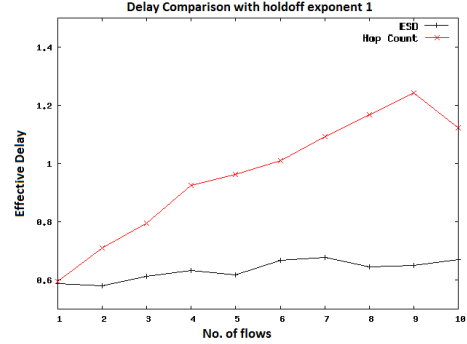

Figure 2. Delay Comparison with HoldOffExp = 1

## IV. PROPOSED SCHEME

In this section we describe our proposed *Expected Scheduler Delay (ESD)* metric for delay sensitive routing, and a distributed source routing algorithm based on modified Bellman-Form algorithm to route the packets in minimum delay path without having any loop.

### A. Proposed Routing Metric - Expected Scheduler Delay (ESD)

Let G (N,E) be a network graph where N is the set of network nodes, and E is the set of edges between the nodes. Let, *X* be an intermediate node in the network. We define the parameter *Flow Descriptor (flowdesc)* at node *X* as follows:

$flowdesc_X$ = total number of incoming and outgoing MAC level flows at node *X*.

Here MAC level flow means a set of packets with the same source and destination MAC addresses.

We assume each node knows the flowdesc of all the nodes in its two hop neighborhoods. Later we describe how it can know this parameter. Now consider two nodes A and B who are directly connected. Now the Expected Scheduler Delay (ESD) from node A to B is defined as follows:

$$ESD = (E[\tau_A] * flowdesc_A + E[\tau_B] * flowdesc_B) / 2 \quad (4)$$

That is the ESD of a link is the average of the expected waiting time at the two ends of the link. The division by 2 ensures that for a multihop path, the delay at any node (except the source and sink) is considered exactly once. Note that the division by 2 is not done for the source and sink. Here we use the parameter flowdesc to avoid congestion at a single path. The parameter actually takes care of the number of flows currently being served by a node. If the IEEE 802.16 mesh uses a Fair Round Robin scheduler, it is justified to multiply the flowdesc with expected waiting time, and then taking the average.

### B. Distributed Mechanism to calculate ESD

In IEEE 802.16 mesh mode scheduling the MSH-DSCH messages are used to carry the distributed scheduling information. In the distributed scheduling mode, each node competes with its 2-hop neighbors for channel access using a pseudo-random election algorithm based on the scheduling information of its 2-hop neighborhood. To inform flowdesc and E[τ] values to all other nodes, each node uses MSH-DSCH messages. Here each node maintains a routing table that contains current flowdesc and E[τ] values of all other nodes. Each node includes this table information in MSH-DSCH message. This can be seen as a modified link state routing, where we broadcast the table information instead of link state table. To reduce the size of the table, each node broadcasts only the updated information, and the time-stamp of last update. Thus once all the node state information is updated, the nodes use the following modified algorithm to find out the minimum delay path to the destination.

### C. Delay Aware Routing Path Computation

A node only uses this modified algorithm when it has the state information of all other nodes. This is a source routing mechanism, where the source finds out the shortest path to the destination, and appends this path information to the packets. Each node uses the Bellman-Ford algorithm to find out the shortest path. To compute the shortest path from Bellman-Ford algorithm, each node constructs the network graph, with the link cost as the ESD. As the value of ESD is non-negative, the routing path is guaranteed to be loop free.

## V. SIMULATION RESULTS

The simulation is done with NS2 [12] simulator. We use the ns2mesh80216-2.33-081113 patch [13] for NS-2.33 to simulate IEEE 802.16 mesh network. The delay and RTT are measured in ms and throughput in bps. Fig.1 shows that our scheme outperforms compared to hop-count specially with high loads. Fig.2 and Fig.3 depict the end-to-end delay comparison with the different HoldOffExp values like 1 and dynamic as proposed in [5]. For a particular constant value of HoldOffExp as the number of flows is increased more packets travel across the network, consequently increasing the flowdesc of the nodes in the network. Hence the delay also increases proportionally with the number of flows. It is seen that as the number of flows increases, the delay of both the metrics increase. For dynamic HoldOffExp values the graph follows an oscillating pattern because the HoldOffExp values vary at different times though the number of flows increases consistently. In all the 3 cases it is observed that the ESD metric gives less end-to-end delay than Hop Count. Thus ESD chooses the path with minimum delay. The MAC delay also reduces the Round-trip-time (RTT), as shown in Fig.4. Fig.5 and Fig.6 depict the performance of ESD with varying grid size. As the graphs imply, our proposed scheme outperforms with varying grid size also. It always finds out the minimum delay path.

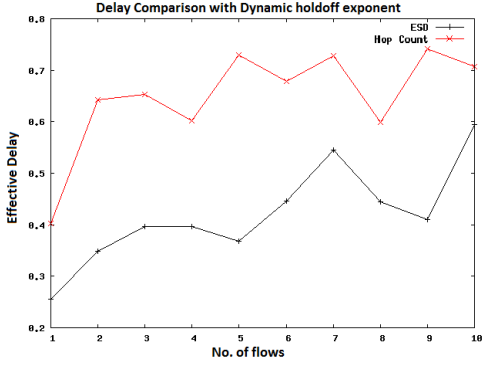

Figure 3. Delay Comparison for Dynamic HoldOffExp

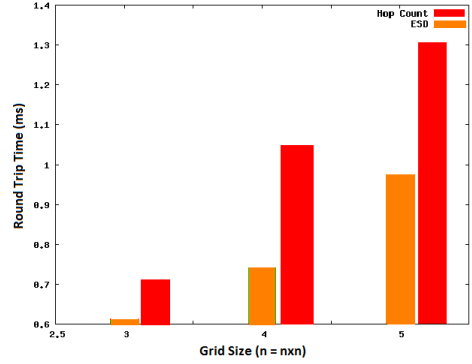

Figure 6. RTT Comparison with varying Grid size

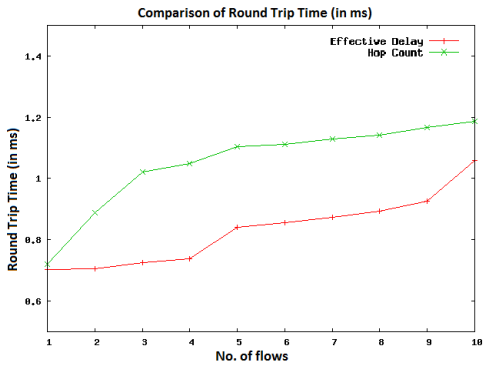

Figure 4. RTT Comparison

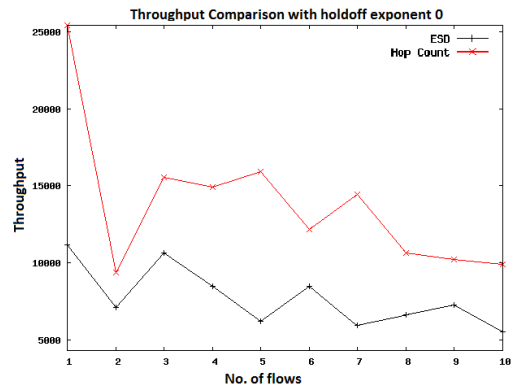

Figure 7. Throughput (in bps) with HoldOffExp = 0

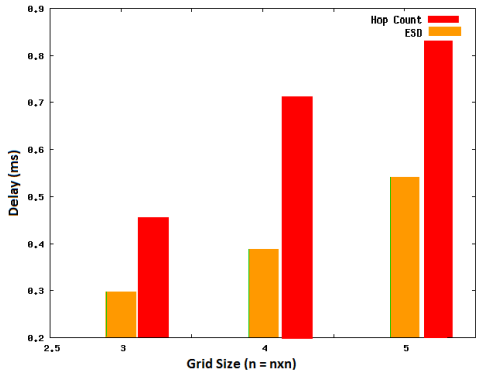

Figure 5. Delay Comparison with varying Grid size

In Fig.7, we have shown a comparison of transport layer end-to-end throughput with HoldOff exponent 0. From the graph, it is clear that in our case, there is a tradeoff between delay and throughput. Here we decrease the delay at the cost of throughput, which can be used in future to design differentiated service architecture between delay sensitive services and best effort services where delay is not a mandatory requirement. This result conforms to the result in [14], that shows the optimal throughput-delay trade-off for multihop wireless network is given by $D(n) = \theta(nT(n))$, where $T(n)$ and $D(n)$ are the throughput and delay respectively.

## VI. CONCLUSION

This paper gives a new cross-layer routing metric *Expected Scheduler Delay (ESD)* which provides an effective route for delay sensitive services in IEEE 802.16 mesh distributed networks. A detailed analysis of this routing metric has been carried out using NS-2 simulations. We find that our new metric is better compared to hop-count metric in terms of delay and RTT. It is expected that this ESD metric will be useful in designing QoS aware routing protocols for IEEE 802.16 mesh distributed networks. The tradeoff between delay and throughput gives a future direction of our research to design differentiated service architecture for delay sensitive services and delay insensitive best effort services.